\documentclass[aps,prd,
preprintnumbers,groupedaddress]{revtex4}
\usepackage{epsbox}
\usepackage{amsmath}
\usepackage{bm}
\usepackage[dvips]{graphicx}

\begin{document}
\preprint{\vtop{{\hbox{YITP-16-51}\vskip-0pt
}}}

\title{
Production of Neutral and Doubly Charged Partners of $\bm{D_{s0}^+(2317)}$ Revisited 
}
\author{Kunihiko Terasaki}
\affiliation{Yukawa Institute for Theoretical Physics, Kyoto University,
Kyoto 606-8502, Japan
}

\thispagestyle{empty}

\begin{abstract}  
Rates for productions of neutral and doubly charged partners of $D_{s0}^+(2317)$ in $B$ 
meson decays are studied by using a hard $D$ meson approximation in the infinite 
momentum frame, and the results are of the same order of magnitude as that of 
$D_{s0}^+(2317)$. Because the bottom-strange $X^{\pm}(5568)$ which can be 
interpreted as iso-triplet bottom partners of $D_{s0}^+(2317)$ have recently been 
discovered, observations of neutral and doubly charged partners of $D_{s0}^+(2317)$ are 
strongly desired. 
\end{abstract}

\maketitle

The charm-strange scalar meson $D_{s0}^+(2317)$ was discovered in the $D_s^+\pi^0$ 
mass distribution in inclusive $e^+e^-$ annihilations~\cite{BABAR-D_spi0,CLEO-D_spi0} 
and also in exclusive $B$ decays~\cite{BELLE-D_spi0}. 
However, no signal of $D_{s0}^+(2317)$ was observed in the $D_s^{*+}\gamma$ channel, 
(its indication was once reported in \cite{BELLE-D_spi0} but nothing after that). 
Thus, the measured value of the ratio of rates  
\begin{equation}
R_{D_{s0}^{+}(2317)}
= \frac
{\Gamma(D_{s0}^{+}(2317) \rightarrow D_{s}^{*+}\gamma)}
         {\Gamma(D_{s0}^{+}(2317) \rightarrow D_{s}^{+}\pi^0)} 
                                                                                     \label{eq:ratio-of-rate}
\end{equation} 
is now given by~\cite{PDG14}  
\begin{equation}
R_{D_{s0}^{+}(2317)}^{\rm exp} < 0.059.                              \label{eq:ratio-of-rate-exp}
\end{equation} 
The above decay property of $D_{s0}^+(2317)$ favors its assignment to an iso-triplet 
state~\cite{Terasaki-D_spi0,HT-isospin}, because there exists a hierarchy of hadron 
interactions, 
$|${\it isospin conserving ones} $\sim O(1)\, |\gg$ 
$|${\it electromagnetic ones} $\sim O(\sqrt{\alpha})\,|\gg$
$|${\it isospin non-conserving ones} $\sim O(\alpha)\,|$~\cite{Dalitz}, 
where $\alpha$ is the fine structure constant. 
In addition, we have seen numerically that decay properties of the charmed vector 
mesons, $D^{*+,0}$ and $D_s^{*+}$, are compatible with the above 
hierarchy~\cite{decay-property}, in particular, the isospin non-conserving 
$D_s^{*+}\rightarrow D_s^+\pi^0$ decay is much weaker than the radiative 
$D_s^{*+}\rightarrow D_s^+\gamma$ as expected by the above hierarchy and 
our theoretical ratio of rates 
$\Gamma(D_s^{*+}\rightarrow D_s^+\pi^0) /\Gamma(D_s^{*+}\rightarrow D_s^+\gamma)$, 
which is dependent on the $\eta\eta'$ mixing angle $\theta_P$,   
is consistent with the measured one~\cite{PDG14}, 
$[\Gamma(D_s^{*+}\rightarrow D_s^+\pi^0) 
   /\Gamma(D_s^{*+}\rightarrow D_s^+\gamma)]_{\rm exp} = 0.062 \pm 0.008$, 
when $\theta_P = -11.4^\circ$~\cite{PDG14} is taken.  
It means that the hierarchy is really working. 
Under this condition, the assignment of $D_{s0}^+(2317)$ to an iso-triplet state implies 
that the $D_{s0}^{+}(2317) \rightarrow D_{s}^{+}\pi^0$ decay is isospin conserving and 
therefore, Eq.~(\ref{eq:ratio-of-rate-exp}) is compatible with the early half of the above 
hierarchy, while its assignment to an iso-singlet state~\cite{CH,BCL} implies that the 
decay is isospin non-conserving, so that Eq.~(\ref{eq:ratio-of-rate-exp}) is against the 
later half of the hierarchy. 
In addition, very recently, iso-triplet bottom-strange $X^\pm(5568)$ have been 
discovered in the $B_s^0\pi^\pm$ channels (based on 10.4 fb$^{-1}$ $p\bar{p}$ collision 
data) by the D0 collaboration~\cite{strange-bottom-D0}, (while the 
LHCb~\cite{strange-bottom-LHCb} have not observed them in 3 fb$^{-1}$ $pp$ collision 
data). 
If such states truly exist, they can be interpreted as the bottom partners of 
$D_{s0}^+(2317)$~\cite{strange-bottom-th}, and it implies that the assignment of 
$D_{s0}^{+}(2317)$ to an iso-triplet state is quite natural. 

When $D_{s0}^{+}(2317)$ is truly an iso-triplet meson, it cannot be realized by any ordinary 
$\{c\bar{s}\}$ state. 
Although one might expect that it could be realized by a $\{DK\}$ molecular state, such 
an expectation would not be realistic, as long as the OZI rule and the isospin $SU_I(2)$ 
symmetry work, as seen below.  
The $\{D^+K^+\}$ and $\{D^0K^0\}$ systems cannot have any $\{q\bar{q}\}$ meson 
exchange in the $t$-channel under the OZI rule~\cite{OZI}, because all the species of 
constituent quarks in each of these systems are different from each other, and 
therefore, no connected $\{q\bar{q}\}$ exchange diagram exists. 
It implies that there exists no ordinary meson exchange as the origin of the binding 
force between $D$ and $K$, as long as the systems are iso-triplet states~\cite{Hyodo}, 
and hence, it is hard to consider that $D_{s0}^{+}(2317)$ is an iso-triplet $\{DK\}$ 
molecular state. 
In this way, we consider that $D_{s0}^{+}(2317)$ is the $I_3 = 0$ component ($\hat{F}_I^+$) 
of iso-triplet charm-strange scalar {\em tetra-quark} mesons, 
$(\hat{F}_I^0,\, \hat{F}_I^+,\, \hat{F}_I^{++})$. 
In this case, however, one might ask how to reconcile a theoretical rate for its isospin 
conserving decay with the measured narrow width of $D_{s0}^{+}(2317)$~\cite{PDG14}. 
Nevertheless, such a problem is not necessarily serious, because tetra-quark states 
have a variety of color and spin configurations, and therefore, the wave function overlap 
between $\hat{F}_I^+$ and $D_s^+\pi^0$ and hence the $\hat{F}_I^+{D}_s^-\pi^0$ coupling 
strength can be suppressed, when the ideally mixed $[qq][\bar{q}\bar{q}],\,(q = u,d,s,c)$ 
are taken as the {\it scalar} tetra-quark mesons~\cite{HT-isospin}. 
A problem is that the Belle collaboration~\cite{search-for-double-charge} have not 
oberved any indication of $\hat{F}_I^0$ and $\hat{F}_I^{++}$ ($z^0$ and $z^{++}$, 
respectively, in \cite{search-for-double-charge}), in spite of our theoretical expectation 
of their existence and observation in $B_u^+$ and $B_d^0$ 
decays~\cite{production-of-double-charge} in addition to the observation of 
$X^\pm(5568)$ as the candidates of iso-triplet bottom partners of $D_{s0}^+(2317)$. 
Therefore, we here re-consider explicitly their production rates in $B$ decays. 

To this aim, we review shortly our previous work on 
productions~\cite{production-of-double-charge} of $D_{s0}^{+}(2317)$ and its partners. 
Productions of  $\hat{F}_I^+ = D_{s0}^+(2317)$ and its iso-singlet partner ($\hat{F}_0^{+}$) 
in inclusive $e^+e^-$ annihilations within the framework of minimal $\{q\bar{q}\}$ pair 
creation is depicted by Fig.~1(a), as in \cite{Scadron-70-KT}. 
In this case, production rate of the iso-triplet $\hat{F}_I^+$ is much higher than that of 
$\hat{F}_0^{+}$, because an iso-triplet $\{n\bar{n}\}$ pair couples more strongly to 
a photon than an iso-singlet one, where $n = u,\,d$, and hence, it is easier to observe 
$\hat{F}_I^+$ than $\hat{F}_0^{+}$ in inclusive $e^+e^-$ annihilations. 
It is consistent with the fact that the Babar and CLEO collaborations have observed 
$\hat{F}_I^+ = D_{s0}^+(2317)$ in the $D_s^+\pi^0$ channel but not any excess
(as the indication of its iso-singlet partner) around the mass of $D_{s0}^+(2317)$ in 
the $D_s^{*+}\gamma$ spectrum~\cite{BABAR-D_spi0,CLEO-D_spi0}. 
In addition, we cannot find any diagram to depict productions of doubly charged and 
neutral partners ($\hat{F}_I^{++}$ and $\hat{F}_I^{0}$) of $D_{s0}^+(2317)$ in inclusive 
$e^+e^-$ annihilations within the same framework, so that we do not expect their 
observation in inclusive $e^+e^-$ annihilations. 
It seems to be consistent with the fact that the Babar collaboration did not observe 
any indication of doubly charged and neutral partners of $D_{s0}^+(2317)$ in inclusive 
$e^+e^-$ annihilations~\cite{BABAR-search}.  

Productions of $\hat{F}_I^+ = D_{s0}^+(2317)$ and $\hat{F}_0^{+}$ in $B$ decays are 
depicted by the quark-line diagrams, Figs.~1(b) and  (e), and its doubly charged and 
neutral partners in $B$ decays by the diagrams, Figs.~1(c) and (d), respectively. 
As seen in these diagrams, amplitudes for these decays are non-factorizable, in contrast 
to those for the $B\rightarrow D_s^+({\rm or}\,\,D_s^{*+})\bar{D}$ decays whose 
amplitudes are factoriable and can be approximately calculated in accordance with 
the vacuum insertion prescription. 
By the way, it is well-known that rates for non-factorizable decays of $B$ mesons are 
much smaller than those for factorizable ones~\cite{Neubert}. 
Therefore, it is easily understood that the measured rates for production of 
$D_{s0}^+(2317)$ in $B_u^+$ and $B_d^0$ decays are much lower than those for  
spectator (factorizable) $B\rightarrow D_s^+({\rm or}\,\, D_s^{*+})\bar{D}$ 
decays~\cite{PDG14}. 
\begin{figure}[t]     %
\includegraphics[width=145mm,clip]{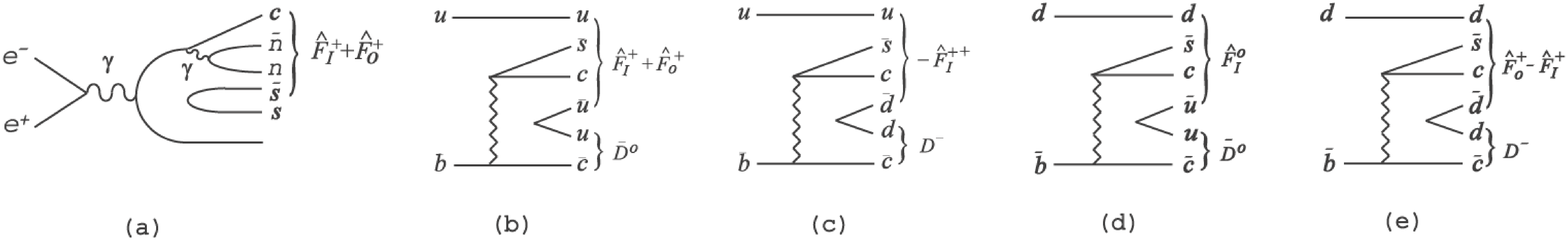}
\label{fig:production-Lisbon.eps}              %
\begin{quote}  
Fig.~1. Productions of charm-strange scalar tetra-quark mesons. 
(a) Productions of $\hat F_I^+$ and $\hat F_0^+$ through electromagnetic 
interactions in inclusive $e^+e^-$ annihilations, (b) and  (e) productions of $\hat F_I^+$ 
and $\hat F_0^+$, and (c) and (d) $\hat{F}_I^{++}$ and $\hat{F}_I^{0}$, respectively, in $B$ 
decays, where $n = u,\,d$. 
\end{quote}
\end{figure}         
Regarding with productions of $\hat F_I^+$ and $\hat{F}_I^{++}$ (as well as $\hat{F}_I^{0}$), 
they are depicted by the same form of diagrams as seen in Fig.~1, so that it is qualitatively 
expected that their rates are not very far from each other, as discussed in our previous 
work~\cite{production-of-double-charge}.
However, this is in contrast to the recent result from the Belle 
experiment~\cite{search-for-double-charge}, as mentioned before. 
Therefore, the decay property of $D_{s0}^+(2317)$ and the productions of its partners 
seems to be inconsistent with each other. 

To see more explicitly this, we study productions of  $\hat{F}_I^+ = D_{s0}^+(2317)$ and 
its partners, $\hat{F}_I^{0}$ and $\hat{F}_I^{++}$, in $B_u^+$ and $B_d^0$ decays, using a 
hard $D$ meson approximation as an extension of a hard pion technique in the infinite 
momentum frame (IMF)~\cite{Oneda-Terasaki-suppl} which can be considered as an 
innovation of the well-known soft-pion technique~\cite{soft-pion}. 
The effective weak Hamiltonian $H_w$ which controlls $B$ decays producing 
charm-strange mesons is provided by  
$H_w = (G_F/\sqrt{2})V_{cs}V_{cb}\bigl\{c_1O_1 + c_2O_2\bigr\} + h.c.$, 
where $O_1$ and $O_2$ are given by 
$O_1 =\,\, :(\bar{c}s)_{V- A}(\bar{b}c)_{V - A}: $ and 
$O_2 =\,\, :(\bar{c}c)_{V- A}(\bar{b}s)_{V - A}:$, 
and $c_1$ and $c_2$ are their coefficients with QCD corrections. 
The CKM matrix elements $V_{ij}$'s~\cite{CKM} are taken to be real, because the $CP$ 
invariance is always assumed in this note. 
It is known that measured rates for spectator (factorizable) decays of charm and 
$B$ mesons are reproduced in a good approximation in the Bauer-Stech-Wirbel (BSW) 
scheme~\cite{Neubert} in which the effective weak Hamiltonian $H_w$ is redefined as, 
$H_w \rightarrow H_w^{\rm BSW} + \tilde{H}_w$, 
where $H_w^{\rm BSW}$ is the so-called BSW Hamiltonian~\cite{BSW} and is given in the 
form  
$H_w^{\rm BSW} = (G_F/\sqrt{2})V_{cs}V_{cb}\bigl\{a_1O_1 + a_2O_2\bigr\} + h.c.$ 
Here, the coefficients of $O_1$ and $O_2$ are given by $a_1 = c_1+ c_2/N_c$ and 
$a_2 = c_2 + c_1/N_c$ with the color degree of freedom $N_c$.  
The terms proportional to $1/N_c$ are extracted from $H_w$ by using the Fierz 
reordering, and the four-quark operators $O_1$ and $O_2$ should be no longer Fierz reordered. 
In the vacuum insertion prescription, matrix elements of $H_w^{\rm BSW}$ taken between 
the initial and final hadron states are factorizable. 
The extra term $\tilde{H}_w$ after extracting a part of $H_w^{\rm BSW}$ (proportional to 
$1/N_c$) from $H_w$ is written as 
$\tilde{H}_w 
= (G_F/\sqrt{2})V_{cs}V_{cb}\bigl\{c_2\tilde{O}_1 + c_1\tilde{O}_2\bigr\} + h.c.$, 
where 
$\tilde{O}_1 = 2\sum_a :(\bar{c}t^a s)_{V - A}(\bar{b}t^a c)_{V - A}:$ and 
$\tilde{O}_2 = 2\sum_a :(\bar{c}t^a c)_{V - A}(\bar{b}t^a s)_{V - A}:$ 
with the generator $t^a$ of the color $SU_c(N_c)$. 
While $\tilde{H}_w$ is taken away under the vacuum insertion prescription, it now survives and provides dynamical contributions of hadrons to hadronic weak decays. 
(For more details, see \cite{non-fact-B,non-fact-K,non-fact-charm}.) 

Now, we study productions of $\hat{F}_I^+ = D_{s0}^+(2317)$ and its partners in $B$ 
decays.  
In the hard $D$ meson approximation in the IMF, it is assumed that the (non-factorizable) 
amplitude for $B(p)\rightarrow \hat{F}(p')\bar{D}(q)$ is approximately given by 
\begin{equation}
M(B\rightarrow \hat{F}\bar{D}) 
\simeq \lim_{\bm{p}\rightarrow \infty,\, \bm{q}\rightarrow 0}M(B\rightarrow \hat{F}\bar{D})
= M_{ETC}(B\rightarrow \hat{F}\bar{D}) + M_S(B\rightarrow \hat{F}\bar{D})      
                                                                                       \label{eq:hard-D-amp}
\end{equation}
under the partially conserved axial-vector current (PCAC) hypothesis, where $B$ 
denotes $B_u^+$ or $B_d^0$, and $\hat{F}$ a charm-strange scalar tetra-quark meson, 
$\hat{F}_I^0$, $\hat{F}_I^+$, $\hat{F}_I^{++}$ or $\hat{F}_0^+$.  
In the above equation, $M_{ETC}$ is the so-called equal-time commutator term which 
is given by 
\begin{equation}
M_{ETC}(B\rightarrow \hat{F}\bar{D}) 
= -\frac{i}{f_D}\langle{\hat{F}|[V_D, \tilde{H}_w(0)]|B}\rangle.    
                                                                               \label{eq:ETC-gen}
\end{equation}
It has the same form as the $ETC$ term in the soft pion approach but it now should be 
evaluated in the IMF. 
$M_S$ is the surface term which is given in the form 
\begin{equation}
M_{S}(B\rightarrow \hat{F}\bar{D}) 
= \lim_{\bm{p}\rightarrow \infty,\, \bm{q}\rightarrow 0}
                                                               \Bigl\{-\frac{i}{f_D}q^\mu T_\mu\Bigr\},
                                                                                          \label{eq:surface-gen}
\end{equation}
where $T_\mu$ is the hypothetical amplitude 
\begin{equation}
T_\mu = i\int e^{iqx} \langle \hat{F}(p')|T[A^{(D)}_\mu (x), \tilde{H}_w(0)]|B(p)\rangle d^4x, 
                                                                                          \label{eq:surface-gen}
\end{equation}
and $A^{(D)}_\mu$ is the axial-vector current with the flavor of $D$. 
$M_{\rm S}$ disappears in the soft pseudoscalar meson ($\pi$, $K$, $\cdots$) 
approximation   but now survives~\cite{Oneda-Terasaki-suppl}, and is given by a sum of 
pole amplitudes, 
\begin{eqnarray}
&&
M_{S}(B\rightarrow \hat{F}\bar{D}) 
= -\frac{i}{f_D}\Bigl\{\sum_{n}\Bigl(\frac{m_{\hat{F}}^2 - m_B^2}{m_n^2 - m_B^2}\Bigr)
\langle{\hat{F}|A_D|n}\rangle\langle{n|\tilde{H}_w|B}\rangle 
- \sum_{\ell}\Bigl(\frac{m_{\hat{F}}^2 - m_B^2}{m_\ell^2 - m_{\hat{F}}^2}\Bigr)
                      \langle{\hat{F}|\tilde{H}_w|\ell}\rangle \langle{\ell|A_D|B}\rangle\Bigr\},   
                                                                                        \label{eq:surface-gen}
\end{eqnarray}
where $f_D$ denotes the decay constant of $D$.  
Here, $V_D$ and $A_D$ are the flavor charge and axial-charge with the flavor of $D$, 
respectively. 
In the above, we have used 
\begin{eqnarray}
&&
\left\{\begin{tabular}{l}
$\displaystyle{\Bigl\{\frac{\langle{\hat{F}(\bm{p}')|V_\pi|n(\bm{p}_n)}\rangle}{N_{\hat{F}}N_n}
\Bigr\}_{\bm{p}_n = \bm{p}\rightarrow \infty}
\hspace{3mm}= (2\pi)^3\delta^{(3)}(\bm{p}_n - \bm{p}')\langle{\hat{F}|V_\pi|n}\rangle}$, \\
$\displaystyle{\Bigl\{\frac{\langle{\hat{F}(\bm{p}')|A_\pi|n(\bm{p}_n)}\rangle}{N_{\hat{F}}N_n}
\Bigr\}_{\bm{p}_n = \bm{p}\rightarrow \infty}
\hspace{2.5mm}
= (2\pi)^3\delta^{(3)}(\bm{p}_n - \bm{p}')\langle{\hat{F}|A_\pi|n}\rangle}$, \vspace{1mm}\\
$\displaystyle{\bigl\{\langle{n(\bm{p}_n)|\tilde{H}|B(\bm{p})}\rangle
                                                           \bigr\}_{\bm{p}_n = \bm{p}\rightarrow \infty} 
\hspace{6mm}= \langle{n|\tilde{H}|B}\rangle
}$
\end{tabular}\right.
\end{eqnarray}
with the normalization factors $N_{\hat{F}}$ and $N_n$ of the state vectors. 
(For more details, see \cite{Oneda-Terasaki-suppl} and 
\cite{non-fact-K,non-fact-charm,non-fact-B}.)
In the intermediate $n$ and $\ell$, single hadron states with the infinite momentum 
survive, and provide the $s$- and $u$-channel poles, respectively. 
Although $M_S$ plays an important role in hadronic weak decays of $K$~\cite{non-fact-K} 
and charm mesons~\cite{non-fact-charm}, their contributions will be not very important 
in $B$ decays, as long as $M_{ETC}$ survives~\cite{hybrid-pict-B}. 
It is because the masses of $B$ mesons are much higher than those of charm and light 
mesons, and therefore, mass dependent factors do not strongly enhance $M_S$ in $B$ 
decays~\cite{non-fact-B}. 
In the present case, the $B\rightarrow \hat{F}\bar{D}$ decay under consideration is 
dominated by a hidden-charm strange tetra-quark state 
$\tilde{\kappa}^c(0^-)\sim \{cn\bar{c}\bar{s}\},\,(n = u,\,d)$ with $J^P = 0^-$ in the 
intermediate $|n\rangle$ and $\langle n|$, and a charmed bottom tetra-quark scalar state 
$\hat{B}_c\sim [nc][\bar{n}\bar{b}],\,(n = u,\,d)$ in the $|\ell\rangle$ and $\langle\ell|$. 
However, in the $s$-channel, the matrix element  
$\langle{\hat{F}|A_D|\tilde{\kappa}^c(0^-)}\rangle$ 
will be small, because the wavefunction overlap between the lowest tetra-quark state 
$\hat{F}$ and the hypothetical $\tilde{\kappa}^c(0^-)$ state with some orbital excitation 
will be small. 
In the $u$-channel, the matrix element of $A_D$ between the hypothetical  
$\langle\hat{B}_c|$ and the initial $|B\rangle$, i.e., the $\bar{B}\hat{{B}}_c\bar{D}$ coupling 
strength, will be small because of a variety of color and spin configurations in the 
tetra-quark state~\cite{HT-isospin}. 
Therefore, contributions of $M_S$ would not be important in decays under consideration, 
because  $M_{ETC}$ survives. 

To evaluate amplitudes under consideration, we parametrize asymptotic matrix elements 
of $V_D$, using the asymptotic flavor $SU_f(4)$ symmetry (the $SU_f(4)$ symmetry of 
matrix elements of $V_D$ taken between single hadron states with the infinite 
momentum)~\cite{Oneda-Terasaki-suppl}. 
We here list their results which will be used later in this note,   
\begin{eqnarray}
&&
\langle \hat{F}_I^{++}|V_{D^+}|\hat{\kappa}^{c+}\rangle 
= \langle \hat{F}_I^{0}|V_{D^0}|\hat{\kappa}^{c0}\rangle
= \sqrt{2}\langle \hat{F}_I^{+}|V_{D^0}|\hat{\kappa}^{c+}\rangle 
= \sqrt{2}\langle \hat{F}_0^{+}|V_{D^0}|\hat{\kappa}^{c+}\rangle 
\nonumber\\
&&
= -\sqrt{2}\langle \hat{F}_I^{+}|V_{D^+}|\hat{\kappa}^{c0}\rangle 
= -\sqrt{2}\langle \hat{F}_0^{+}|V_{D^+}|\hat{\kappa}^{c0}\rangle  
= \cdots = \langle{\hat{F}_I^{++}|V_{\pi^0}|\hat{F}_I^{++}}\rangle = 1,  
                                                                          \label{eq:matrix-elements-V_D}
\end{eqnarray}
where $\hat{\kappa}^c$'s denote hidden-charm strange members of scalar tetra-quark 
mesons in our model~\cite{Terasaki-D_spi0}, i.e., 
$\hat{\kappa}^c \sim [cn][\bar{c}\bar{s}],\,(n = u,\,d)$. 
Here, it should be noted that the above parametrization might cause about $20 -30$ per 
cent errors in amplitudes. 
It is because matrix elements of $V_D$ taken between single hadron states are given by a 
form factor of a charm changing vector current at zero momentum tranaser squared 
which is normalized to be unity in the $SU_f(4)$ symmetry limit, while the measured form 
factors have been given as~\cite{PDG96} 
\begin{equation}
f_+^{(\bar{K}D)}(0) = 0.74 \pm 0.03\quad {\rm and}\quad 
\frac{f_+^{(\bar{\pi}D)}(0)}{f_+^{(\bar{K}D)}(0)} = 
\left\{\begin{tabular}{l}
$1.00 \pm 0.11\pm 0.02$\hspace{4mm} (Fermilab E687), \\
$0.99\pm 0.08$\hspace{15mm} (CLEO), 
\end{tabular}\right.
\label{eq:FF-KD-exp}
\end{equation}
where the results from Fermilab E687 and CLEO have been given in \cite{FNAL-E687} 
and \cite{CLEO}, respectively. 
Regarding with a parametrization of asymptotic matrix elements of $A_D$, however, we 
skip to list their results, because we have neglected contributions of $M_S$ which 
contains matrix elements of axial-charges. 
In this way, we here list amplitudes for productions of charm-strange scalar tetra-quark 
mesons in $B$ decays as  
\begin{eqnarray}
&& \hspace{-10mm} 
M(B_u^+\rightarrow \hat{F}_I^{++}D^-) 
\simeq 
\hspace{8.5mm}-\frac{i}{f_D}\Bigl\{\langle{\hat{\kappa}^{c+}|\tilde{H}_w|B_u^+}\rangle \,\,
+ \cdots\Bigr\}, 
                                                                     \label{eq:B_u-double-charge-full}
\\
&& \hspace{-10mm} 
M(B_d^0\rightarrow \hat{F}_I^{0}\bar{D}^0) \hspace{4.5mm}
\simeq \hspace{8.3mm}-\frac{i}{f_D}\Bigl\{
\,\langle{\hat{\kappa}^{c0}|\tilde{H}_w|B_d^0}\rangle\,\,\,\,
+ \cdots\Bigr\},  
                                                                              \label{eq:B_d-neutral-full}
\\
&&\hspace{-10mm}
M(B_u^+\rightarrow \hat{F}_I^{+}\bar{D}^0) \,\,\,\,\,
\simeq \,\,\,\,-\sqrt{\frac{1}{2}}\frac{i}{f_D}\Bigl\{
                                        \langle{\hat{\kappa}^{c+}|\tilde{H}_w|B_u^+}\rangle \,\,
+ \cdots \Bigr\},  
                                                                         \label{eq:B_u-single-charge-full}
\\
&&\hspace{-10mm}
M(B_u^+\rightarrow \hat{F}_0^{+}\bar{D}^0) \hspace{2mm}\,\,
\simeq \hspace{2mm}-\sqrt{\frac{1}{2}}\frac{i}{f_D}\Bigl\{
                                        \langle{\hat{\kappa}^{c+}|\tilde{H}_w|B_u^+}\rangle \,\,
+ \cdots \Bigr\},  
                                                                                   \label{eq:B_u-singlet-full}
\\
&&\hspace{-10mm}
M(B_d^0\rightarrow \hat{F}_I^+D^-) \,\,\,\,\,
\simeq \hspace{5mm}\sqrt{\frac{1}{2}}\frac{i}{f_D}\Bigl\{\,
\langle{\hat{\kappa}^{c0}|\tilde{H}_w|B_d^0}\rangle \,\,\,
+ \cdots \Bigr\}, 
                                                                          \label{eq:B_d-single-charge-full}
\\
&& \hspace{-10mm} 
M(B_d^0\rightarrow \hat{F}_0^{+}{D}^-) \hspace{3mm}
\simeq \hspace{5mm}\sqrt{\frac{1}{2}}\frac{i}{f_D}\Bigl\{\,
\langle{\hat{\kappa}^{c0}|\tilde{H}_w|B_d^0}\rangle \,\,\,
+ \cdots \Bigr\},
                                                                          \label{eq:B_d-singlet-full}
\end{eqnarray}
in the approximation in which $M_S$ is neglected, where the ellipses denote the 
neglected contributions of $M_S$. 
As seen in Eqs.~(\ref{eq:B_u-double-charge-full}) -- (\ref{eq:B_d-singlet-full}), these 
amplitudes are given by 
$\langle{\hat{\kappa}^{c+}|\tilde{H}_w|B_u^+}\rangle$ 
or $\langle{\hat{\kappa}^{c0}|\tilde{H}_w|B_d^0}\rangle$. 
Difference between the above two matrix elements of $\tilde{H}_w$ is in their spectator 
quarks, i.e., $u$ in the former and $d$ in the latter. 
Therefore, these two matrix elements are equivalent to each other in the $SU_I(2)$ 
symmetry limit. 
With these approximations, we obtain the following relations of rates for productions of 
charm-strange scalar tetra-quark mesons,  
\begin{eqnarray}
&& \hspace{-10mm} 
\Gamma(B_u^+\rightarrow \hat{F}_I^{++}D^-) 
\simeq 2\Gamma(B_u^+\rightarrow \hat{F}_0^{+}\bar{D}^0) 
\simeq 2\Gamma(B_u^+\rightarrow \hat{F}_I^{+}\bar{D}^0) 
\nonumber\\
&& \hspace{-9.5mm} 
\simeq \Gamma(B_d^0\rightarrow \hat{F}_I^{0}\bar{D}^0) 
\simeq 2\Gamma(B_d^0\rightarrow \hat{F}_0^{+}{D}^-)  
\simeq 2\Gamma(B_d^0\rightarrow \hat{F}_I^{+}{D}^-), 
                                                                                \label{eq:relations-of-rates}
\end{eqnarray}
when masses of $\hat{F}_0^+$, $\hat{F}_I^0$, $\hat{F}_I^+$ and $\hat{F}_I^{++}$ are 
approximately degenerate. 

Because the measured lifetimes of $B_u^+$ and $B_d^0$ are not equal to each 
other~\cite{PDG14}, 
\begin{eqnarray}
&& \hspace{-10mm} 
(\tau_{B_u^+})_{\rm exp} = (1.641\pm 0.008)\times 10^{-12}\,\, s\quad{\rm and}\quad
(\tau_{B_d^0})_{\rm exp} = (1.519 \pm 0.007)\times 10^{-12} \,\,s, 
 \label{eq:lifetime-of-B}
 \end{eqnarray}
however, we obtain separately 
\begin{eqnarray}
&& \hspace{-10mm} 
\mathcal{B}(B_u^+\rightarrow \hat{F}_I^{++}D^-) 
\simeq 2\mathcal{B}(B_u^+\rightarrow \hat{F}_0^{+}\bar{D}^0) 
\simeq 2\mathcal{B}(B_u^+\rightarrow \hat{F}_I^{+}\bar{D}^0) 
= 2\times \bigl(7.3^{+2.2}_{-1.7}\bigr)\times 10^{-4}
                                                                                     \label{eq:br-of-B_u-th}
\end{eqnarray}
and 
\begin{eqnarray}
&& \hspace{-10mm} 
\mathcal{B}(B_d^0\rightarrow \hat{F}_I^{0}\bar{D}^0) 
\simeq 2\mathcal{B}(B_d^0\rightarrow \hat{F}_0^{+}{D}^-) 
\simeq 2\mathcal{B}(B_d^0\rightarrow \hat{F}_I^{+}{D}^-) 
= 2\times \bigl(9.7^{+4.0}_{-3.3}\bigr)\times 10^{-4},  
                                                                                    \label{eq:br-of-B_d-th}
\end{eqnarray}
where their measured branching fractions~\cite{PDG14} have been taken in the above.  
Although the measured 
$\mathcal{B}(B_u^+\rightarrow \hat{F}_I^{+}\bar{D}^0)$ and 
$\mathcal{B}(B_d^0\rightarrow \hat{F}_I^{+}{D}^-)$ 
in Eqs.~(\ref{eq:br-of-B_u-th}) and (\ref{eq:br-of-B_d-th}) still have large uncertainties, 
they are compatible with our approximate equality, 
$\Gamma(B_u^+\rightarrow \hat{F}_I^{+}\bar{D}^0) 
\simeq \Gamma(B_d^0\rightarrow \hat{F}_I^{+}{D}^-)$, 
in Eq.~(\ref{eq:relations-of-rates}). 
This seems to mean that our approach is not so far from the reality. 
However, the same equation predicts that rates for $\hat{F}_I^{++}$ and $\hat{F}_I^{0}$ 
productions are of the same order of magnitude as (larger by about a factor two than) 
those of $\hat{F}_I^{+}$, and also that rates for productions of the iso-singlet partner 
$\hat{F}_0^+$ in $B_u^+$ and $B_d^0$ decays are approximately equal to those of 
$\hat{F}_I^+$ productions, as expected qualitatively in our previous 
work~\cite{production-of-double-charge}. 
In addition, if a peak around the mass of $D_{s0}^+(2317)$ in the $D_s^{*+}\gamma$ mass 
spectrum is observed, it would be an indication of $\hat{F}_0^+$, because it should decay 
dominantly into $D_s^{*+}\gamma$ as discussed before. 
Therefore, the fact that experiments did not observe any signal of the 
$D_{s0}^+(2317)\rightarrow D_s^{*+}\gamma$ decay as seen in 
Eq.~(\ref{eq:ratio-of-rate-exp}) (i.e., $D_{s0}^+(2317)$ favors its assignment into an 
iso-triplet state) is not compatible with the fact that the Belle collaboration did not 
observe any signal of $\hat{F}_I^{++} = z^{++}$ and $\hat{F}_I^{0} = z^0$ in $B$ decays. 
Nevertheless, observation of $D_{s0}^{0}(2317)$ and $D_{s0}^{++}(2317)$ in $B$ decays 
are strongly desired, because the recently observed $X^\pm(5568)$ can be interpreted 
as the iso-triplet bottom partners of $D_{s0}^+(2317)$.  

So far, we have considered that $D_{s0}^+(2317)$ is an iso-triplet tetra-quark scalar 
meson. 
However, there exists an argument~\cite{Faessler} that the rate for the isospin 
non-conserving $D_{s0}^{+}(2317)\rightarrow D_s^+\pi^0$ decay can overcome that for 
the radiative $D_{s0}^{+}(2317)\rightarrow D_s^{*+}\gamma$ against the hierarchy of 
hadron interactions, if $D_{s0}^+(2317)$ is a $\{DK\}$ molecule (even if it is an iso-singlet 
state), i.e., the intermediate $DK$ loop contributions enhance extraordinarily the amplitude 
for the isospin non-conserving $D_{s0}^{+}(2317)\rightarrow D_s^+\pi^0$ decay, and as 
the result, this model leads to a ratio of decay rates compatible with the measured one, 
Eq.~(\ref{eq:ratio-of-rate-exp}). 
By the way, in order that this model works well, it seems to be implicitly required that 
the constituent $D$ and $K$ are sufficiently compact (or local). 
In the above analysis, however, a size parameter $\Lambda_{D_{s0}(2317)}$ which 
parametrizes the distribution of the constituent $D$ and $K$ in $D^+_{s0}(2317)$ has 
been introduced, and the cases with $\Lambda_{D_{s0}(2317)} = 1 - 2$ GeV (compact) and 
$\infty$ (the local limit) have been investigated. 
This implies that $\Lambda_{D_{s0}(2317)}^{-1}$, which can be very crudely considered as 
the size of the $\{DK\}$ molecule, is smaller than the measured size of the constituent 
$K$ meson, as seen below. 
It is considered that the size of $K$ meson is approximately given by the charge radius 
of $K^\pm$, which can be determined by measurements of the 
$eK^\pm\rightarrow eK^\pm$ scattering. 
A typical result on the mean square charge radius of $K^\pm$ has been provided 
as~\cite{CERN-FF-K} $\langle r^2\rangle_{K^\pm} = (0.34 \pm 0.05)\,{\rm fm}^2$, so that 
the charge radius is $\sqrt{\langle r^2\rangle_{K^\pm}}\simeq 0.58\,{\rm fm}$.  
This result is considerably larger than the maximum value  
$(\Lambda_{D_{s0}(2317)}^{-1})_{\rm max} = 1\,{\rm GeV}^{-1}$ 
considered in \cite{Faessler}. 
This implies that the constituent $K$ meson in the $\{DK\}$ molecule under 
consideration is not sufficiently compact. 
Therefore, the $\{DK\}$ molecular picture of ${D}_{s0}^+(2317)$ in \cite{Faessler} seems 
to be unrealistic.  

In summary we have discussed that the decay property of $D_{s0}^+(2317)$ favors its 
assignment to an iso-triplet scalar state, because of the hierarchy of hadron interactions. 
In this case, it is expected that there exist its neutral and doubly charged partners, 
$D_{s0}^0(2317) = \hat{F}_I^0$ and $D_{s0}^{++}(2317) = \hat{F}_I^{++}$. 
As seen in the quark-line diagrams, amplitudes for their productions, 
$B\rightarrow \hat{F}\bar{D}$'s, are non-factorizable, and therefore, their rates have 
been calculated by using the hard $D$ meson approximation in the IMF, where $\hat{F}$ 
denotes $\hat{F}_I^0$, $\hat{F}_I^+$, $\hat{F}_I^{++}$ or $\hat{F}_0^+$. 
As the result, we have seen that the expected rates for productions of $\hat{F}_I^{++}$ 
and $\hat{F}_I^{0}$ are approximately equal to each other, and that they are of the same 
order of magnitude as that for the productions of $\hat{F}_I^+ = D_{s0}^+(2317)$ in 
$B_u^+$ and $B_d^0$ decays, in contrast to the result from the Belle collaboration that 
no signal of $\hat{F}_I^{++}$ and $\hat{F}_I^{0}$ was observed in $B$ decays. 
In addition, we have provided a brief comment on an iso-singlet $\{DK\}$ molecular 
model of $D_{s0}^+(2317)$ which leads to a ratio of rates consistent with  
the measured one, Eq.~(\ref{eq:ratio-of-rate-exp}).   
That is, the constituent $K$ meson seems to be not sufficiently compact in this model, 
so that this model seems to be unnatural. 
In this way, it has been discussed that the decay property of $D_{s0}^+(2317)$ and 
the productions of its neutral and doubly charged partners are not compatible with 
each other. 
Because $X^\pm(5568)$ which can be interpreted as the iso-triplet bottom partners of 
$D_{s0}^+(2317)$ have been discovered, however, it is strongly desired that experiments 
will observe $D_{s0}^{0}(2317)$ and $D_{s0}^{++}(2317)$. 

\section*{Acknowledgments} 
The author would like to thank Prof.~T.~Hyodo, Yukawa Institute for Theoretical 
Physics, Kyoto University for valuable discussions and comments. 
He also would like to appreciate Prof.~H.~Kunitomo, Yukawa Institute for Theoretical 
Physics, Kyoto University for careful reading of the manuscript. 


\end{document}